\documentclass{ws-procs975x65}
\usepackage{euscript}
\def\beq{\begin{equation}}
\def\eeq{\end{equation}}
\def\beq{\begin{equation}}
\def\eeq{\end{equation}}

\begin{document}

\title{Geometric approaches to the thermodynamics of black holes}
\author{Christine Gruber$^*$}

\address{Instituto de Ciencias Nucleares, Universidad Nacional Aut\'{o}noma de M\'{e}xico, \\
AP 70543, M\'{e}xico, DF 04510, M\'{e}xico \\
$^*$E-mail: christine.gruber@correo.nucleares.unam.mx}

\author{Orlando Luongo$^{\dagger}$}

\address{Department of Mathematics and Applied Mathematics, University of Cape Town, Rondebosch 7701, Cape Town, South Africa.\\
Astrophysics, Cosmology and Gravity Centre (ACGC), University of Cape Town, Rondebosch 7701, Cape Town, South Africa.\\
Dipartimento di Fisica, Universit\`a di Napoli ''Federico II'', Via Cinthia, Napoli, Italy.\\
Istituto Nazionale di Fisica Nucleare (INFN), Sez. di Napoli, Via Cinthia, Napoli, Italy.\\
$^{\dagger}$E-mail: luongo@na.infn.it, luongo@uct.ac.za}

\author{Hernando Quevedo$^{\ddagger}$}

\address{Instituto de Ciencias Nucleares, Universidad Nacional Aut\'{o}noma de M\'{e}xico, \\
AP 70543, M\'{e}xico, DF 04510, M\'{e}xico \\
$^{\ddagger}$E-mail: quevedo@nucleares.unam.mx}

\begin{abstract}

In this summary, we present the main topics of the talks presented in the parallel session 
``Black holes - 5'' of the 14th Marcel Grossmann Meeting held in Rome, Italy in July 2015.
We first present a short review of the main approaches used to understand thermodynamics 
by using differential geometry. Then, we present a brief summary of each presentation,
including some general remarks and comments.   

\end{abstract}


\bodymatter

\section{Introduction}

Differential geometry is a very important tool of mathematical physics with 
diverse applications in physics,
chemistry, engineering and even economics. As one of the most important applications from
the point of view of theoretical physics,  we can mention the case
of the four interactions of Nature for which a well-established description in terms of geometrical concepts
is known. 
Indeed, Einstein proposed in 1915 the astonishing principle ''field strength = curvature`` 
to understand the physics of the gravitational field \cite{frankel}. 
In an attempt to associate a geometric structure to  the electromagnetic field, Yang and Mills \cite{ym53}
used in 1953 the concept of a principal fiber bundle with the Minkowski spacetime, as the base manifold, and the symmetry group $U(1)$,
as the standard fiber, to demonstrate that the Faraday tensor can be interpreted as the curvature of this particular fiber bundle.
Today, it is well known \cite{frankel} that the weak interaction and the strong interaction can be represented as the 
curvature of a principal fiber bundle with a Minkowski base manifold and the standard fiber $SU(2)$ and $SU(3)$, respectively.

In very broad terms, one can say that all the known forces of Nature act among the particles that constitute a thermodynamic system.  
Due to the large number of particles involved in the system, only a statistical approach is possible, from which average thermodynamic 
values for the physical quantities of interest are derived. 
Although the  thermodynamic laws are  based entirely upon empirical results which are satisfied under certain
conditions in almost any macroscopic system, the geometric approach to thermodynamics
has proved to be very useful. In our opinion, the following three branches of geometry
have found sound applications in equilibrium thermodynamics: analytic geometry, Riemannian geometry, and contact geometry.

One of the most important contributions of analytic geometry
to the understanding of thermodynamics is the identification of points of phase transitions with
extremal points of the surface determined by the state equation of the corresponding thermodynamic system. 
For a more detailed description of these contributions see, for instance, the books by H. B. Callen \cite{callen} or 
K. Huang \cite{huang}.

Riemannian geometry was first introduced in statistical physics and thermodynamics
by Rao \cite{rao45}, in 1945.  Rao introduced a metric whose components in local coordinates coincide with Fisher's
information matrix. Rao's pioneering  work has been followed up and extended by a number of
authors (for a review, see, e.g. the book by S. Amari \cite{amari85}). On the other hand, Riemannian geometry in the space of
equilibrium states was introduced by Weinhold \cite{wei75,wei75-2} and Ruppeiner \cite{rupp1979,rup95,rup96},
who defined metric structures as the Hessian of the internal energy and the entropy, respectively.
Both metrics have been widely used to study the geometry of the equilibrium space of
ordinary systems and black holes. The approach based upon the use of Hessian metrics is commonly known 
as thermodynamic geometry.

Contact geometry was introduced by Hermann \cite{her73} into the thermodynamic phase space
in order to formulate in a consistent manner the geometric version of the laws of thermodynamics. 
One important property of classical thermodynamics is that it does not depend on the choice of 
thermodynamic potential which is equivalent to saying that it is invariant with respect to 
Legendre transformations \cite{callen}. Contact geometry allows us to consider a Legendre 
transformation as a coordinate transformation in the phase space \cite{arnold}. This fact
was used by Quevedo \cite{quev07} to propose the formalism of geometrothermodynamics which
takes into account the Legendre invariance of classical thermodynamics, and is considerably different 
from the approach of thermodynamic geometry.

One of the goals of this parallel session of MG14 was to allow researchers interested in the relationship
between geometry and thermodynamics to present their results and applications, 
especially in the context of black hole
thermodynamics. This goal was reached to some extent. All the talks discussed problems related 
to black hole thermodynamics in different theories, its physical implications, phase transition 
structures, equilibrium spaces, thermodynamic phase spaces and others. 

This work is split into several sections, each of which corresponds to one presentation in the parallel 
session BH5. 
In this way, we hope to take into account the interests of all the speakers, and reflect the spirit of the session
held at a highly scientific level and carried out in full cordiality.

\section{Possible Observation Sequences Randall-Sundrum Black Holes}
In the first talk of the session, Erin Nikita presented his work on sperically symmetric Randall-Sundrum (RS) black 
holes, done in collaboration with Kristina A. Rannu and Stanislav O. Alexeyev. They investigated the ACPY-solution \cite{2013Abdo}, 
which is an asymptotically Schwarzschild solution of $5$-dimensional RS models, given by the metric 
\begin{eqnarray}
  ds^2 &=  \left[ 1- \cfrac{1}{-\Lambda_5 r^2} \cfrac{r-2M}{2-1.5M} \left( F - r \cfrac{dF}{dr} \right) \right] 
      \left( 1 - \cfrac{2M}{r} \right)^{-1} \ dr^2 \nonumber\\ 
  & ~~~~~~~~~~~~ - \left( 1 - \cfrac{2M}{r} \right) dt^2 + \left[ r^2 + \cfrac{F}{- \Lambda_5} \right] d \Omega^2 \,. 
\end{eqnarray}
Here, $\Lambda_5$ is a negative cosmological constant in $5$ dimensions, and $F$ is a numerically found polynomial of 
$11$\textsuperscript{th} order, 
\begin{eqnarray} 
    & F(r) = 1 - 1.1241 \left( \cfrac{2M}{r} \right) + 1.956 \left( \cfrac{2M}{r} \right)^2 \\
    & ~~~~~~~~~~~~~~~~~~~~~~~~~~ - 9.961 \left( \cfrac{2M}{r} \right)^{3} + \ \dots \ + 2.900 \left( \cfrac{2M}{r} \right)^{11} \,.
\end{eqnarray}
Even in the post-Newtonian regime RS models can hardly be distinguished from Einstein gravity. A possibility to derive 
observational properties which could be checked against observations is to consider accretion disks around and the thermodynamic 
properties of the black hole solution. Concretely, the lifetime of primordial black holes in the RS frameworks, nucleated in 
the early universe at high temperatures and density fluctuations, have been considered. Initial masses of these primordial black 
holes which reach the end of their lifetime around the present time are assumed to be of the order of 
\begin{equation} 
  M_{\rm 0} \approx 5.0 \times 10^{14} \ \mbox{g} \,. 
\end{equation}
Their radiation contributes to the cosmic microwave background \cite{1997Frol}, and in the final stages of their evaporation 
they are expected to produce bursts of energy \cite{2012Alex}, predicted to be gamma radiation in the range of MeV to TeV, at redshifts 
of $z \leq 9.4$ \cite{2011Cucc}. 

Present time telescopes are capable of detecting the evaporation of PBHs at maximal distances $d$ of 
\begin{equation} 
  d \simeq 0.04 \left( \frac{\Omega}{\mbox{sr}} \right)^{-0.5} \left(
    \frac{E}{\mbox{GeV}} \right)^{0.7} \left( \frac{T}{\mbox{TeV}} \right)^{0.8} \mbox{pc} \,,
\end{equation}
where $\Omega$ is the angular telescope resolution, $T$ is the black hole temperature and $E$ represents the energy range of the 
telescope. 

For detection of differences between GR and RS models in the evaporation of primordial black holes, the difference in masses is 
supposed to be \cite{2015Alex} 
\begin{equation} 
  \cfrac{M_{RS}}{M_{GR}} > 10^5 \,.
\end{equation}
The calculations of black hole evaporation, following the standard expression by Hawking \cite{1974Hawk}, for the ACPY solution 
provides a precise expression for the evaporation rate in terms of the mass up to $10$\textsuperscript{th} order, 
\begin{equation}
  -\frac{dM}{dt} \simeq \frac{1}{256} \frac{k_B}{\pi^3 M^2} + \frac{1}{512} \frac{k_B}{\pi^3 M^6} + \mathcal{O} \left( M^{-10} \right) \,.
\end{equation}
This result shows that the present conditions and gamma ray bursts of currently evaporating primordial black holes are not enough 
to distinguish standard Einstein gravity black holes from RS black holes.

\section{Constructing black hole entropy from gravitational collapse}
A new perspective to describe thermodynamic properties of black hole formation has been discussed by Aymen Hamid, developed in 
collaboration with Giovanni Acquaviva, George F. R. Ellis and Rituparno Goswami \cite{2015Acqu}, through the use of gravitational 
collapse in semitetrad 1+1+2 covariant formalism. This approach extends the 1+3 formalism to any spacetime with a preferred 
spatial direction. Particular attention is here devoted to the simplest case of spherically symmetric metrics. 

Under the simplest hypothesis of Oppenheimer-Snyder-Datt collapse, a spherical dust like star living on Schwarzschild geometry 
was assumed, the star's exterior classified as Petrov type D. The entropy of the free gravitational field can be inferred for a 
static observer, even even if no event horizons exist. Considering a notion of entropy during the collapse process allows for 
an answer to the question whether entropy in this context is a property of the horizon only, which emerges after the creation 
of a black hole, or whether it can in principle be defined at all times and smoothly changes from an initial configuration to 
its black hole result $S_{BH} = A/4$. 

The gravitational entropy will be denoted as $S_{grav}$ and its definition is based on the Bel-Robinson tensor \cite{1958Bel}. 
In order to be compatible with standard features of physical entropy, it is necessary to demand that the entropy should be a 
measure of the local anisotropy of the free gravitational field, non-negative, vanishing for zero Weyl tensor, and in the 
limiting case should result in the Bekenstein-Hawking entropy of black holes. 

From those requirements, assuming that the second law of thermodynamics still continues to hold, the relation 
\begin{equation}\label{seconda}
 T_{grav}dS_{grav}=dU_{grav}+p_{grav}dV>0
\end{equation}
has to be valid, with $T_{grav}$, $U_{grav}$ and $p_{grav}$ the effective temperature, internal energy and isotropic pressure of 
the free gravitational field, and $V$ the spatial volume. With the gravitational pressure vanishing in a Coulomb-like field, 
$p_{grav}=0$, and employing the equation of energy conservation, the second law and also the temperature of the gravitational 
field can be expressed in terms of the kinematical and Weyl quantities of the spacetime. Calculating the change of the entropy 
of the gravitational field during a time interval $(\tau - \tau_0)$ then yields the result 
\begin{equation}
  \delta S_{grav}|_{(\tau-\tau_0)} = \frac{\alpha}{4}\Big(A(\tau_0)-A(\tau)\Big) \,, 
  \label{deltas}
\end{equation}
where $A(\tau)$ is the surface area of the star at an arbitrary time $\tau>\tau_0$, and $\alpha$ is a constant introduced in the 
definition of the gravitational energy-momentum tensor, which can be constrained to the value $\alpha=1$. 

The increase in the instantaneous gravitational entropy outside a collapsing star during a given interval of time is thus proportional 
to the change in the surface area of the star during that interval, and thus the black hole form of the entropy is recovered in the 
process of gravitational collapse.

\section{A Heuristic Energy Quantization Of Equilibrium Black Hole Horizons}
In his contribution, Abhishek Majhi aimed at deriving the energy quantization of black holes 
by employing heuristic arguments from thermodynamics and statistical mechanics \cite{2013Mahj}. 

The black holes are hereby considered as isolated horizons, i.e., 3-dimensional null inner boundaries of a 4-dimensional 
spacetime, and possess topology $S(2) \times R$. The symplectic structure of the 4-d bulk spacetime induces a symplectic 
structure of a $SU(2)$ Chern-Simons (CS) theory on the boundary, i.e., the isolated horizon. The quantization of the isolated 
horizon in a straightforward manner is not possible, since Chern-Simons theories are topological field theories in which 
the Hamiltonian vanishes identically, and thus no well-defined notion of energy exists. However, naturally the bulk 
spacetime enclosed by the boundary entails a notion of energy, and thus there is a first law of thermodynamics associated 
with the boundary, which in turn assigns a meaning of energy to the boundary field theory. Moreover, a correspondence of 
bulk and boundary states can be achieved in Loop Quantum gravity (LQG), where states are represented by sets of spins 
$|j_1, ..., j_N \rangle$. A general Hamiltonian should act on states as 
\begin{equation}
  \hat{H}_S |j_1, \dots, j_N \rangle = l_p \sum_{l=1}^{N} \epsilon_{jl} | j_1, ..., j_N \rangle \,,
\end{equation}
with the energy eigenvalues $\epsilon_{jl}$. 

It is possible to use these states and the associated Hilbert space to count the possible microstates on the isolated 
horizon, i.e., the entropy, under the assumption of constant area, which corresponds to the constant entropy ensemble. 
Using an approach from LQG, the spectrum of the spins representing the state of the system is calculated, and the distribution 
function for the most probable configuration of spins is obtained from extremizing the entropy. The entropy of the isolated 
horizon can then be written as 
\begin{equation}
  S = \frac{\lambda_0 A_{IH} }{8 \pi \gamma l_p^2} \,,
\end{equation}
where $A_{IH}$ is the classical area of the isolated horizon, $\gamma$ is the Barbero-Immirzi parameter and $l_p^2$ is 
the Planck length. A Lagrange multiplier $\lambda$ has been introduced in the extremization procedure, taking on the value 
of $\lambda_0$ for the most probable spin configuration. 

The constant entropy approach is however only one possible ensemble in which the system can be considered. The physical 
properties of the system should not depend on the choice of ensemble, and therefore the treatment in another ensemble, 
e.g. one with constant energy, must be completely equivalent and yield the same physical predictions. Extremizing the 
entropy under the constraint of constant energy, using a Lagrange multiplier $\beta$, the energy and entropy of the 
isolated horizon are then related as 
\begin{equation}
  S = \frac{\beta E_{IH}}{l_p} \,.
\end{equation}
With this expression, $\beta$ can then be interpreted as the inverse temperature of the horizon. 

Comparing the obtained relation of entropy/area and energy, 
\begin{equation}
  \delta E_{IH} = \frac{\lambda_0}{8\pi \gamma \beta l_p} \delta A_{IH} 
\end{equation}
with the first law of thermodynamics, 
\begin{equation}
  \delta E_{IH} = \frac{\kappa_{IH}}{8\pi} \delta A_{IH} \,,
\end{equation}
where $\kappa_{IH}$ is the surface gravity of the horizon, it is possible to identify 
\begin{equation}
  \frac{1}{\beta} = \frac{\gamma l_p}{\lambda_0} \kappa_{IH} \,.
\end{equation}
From the general formulation of the Hamiltonian of the system, and using this relation between the constants, a 
Hamiltonian is then defined as 
\begin{equation}
  \hat{H}_S |j_1, \dots, j_N \rangle = \frac{\kappa_{IH}}{8\pi} |j_1, ..., j_N \rangle \,.
\end{equation}
The Hamiltonian captures the physics associated with both near horizon Rindler observers as well as asymptotic observers, 
and allows for the formulation of black hole thermodynamics in the usual energy ensemble. This work thus bypasses the problem 
of the vanishing Hamiltonian in the boundary Chern-Simons theory with arguments from thermodynamics and statistical mechanics.

\section{Thermodynamic Volume And Phase Transitions Of AdS Black Holes}
In his talk, David Kubiznak presented a new point of view on black hole thermodynamics, in particular on black holes with 
a cosmological constant. If the cosmological constant is identified with pressure, and its conjugate quantity with thermodynamic 
volume, black holes can be understood from a chemical perspective in terms of concepts such as van der Waals fluids, reentrant 
phase transitions, and triple points. In a series of articles in collaboration with Robert Mann, Natacha Altamirano and Zeinab 
Sherkatghanad, \cite{2012Kubi,2013Alta,2014Alta1,2014Alta2,2015Kubi}, various types of AdS black holes, including rotating (around 
one or more axes), charged and higher dimensional ones, have been investigated. 

The principal idea is to treat the cosmological constant as a thermodynamic variable, i.e., the pressure, including it into the 
first law of thermodynamics. Identifying the pressure through comparison of the Smarr relation with Euler's theorem for homogeneous 
functions as 
\begin{equation}
  P = -\frac{\Lambda}{8\pi} = \frac{(d-1)(d-2)}{8\pi l^2} \,,
\end{equation}
where $\Lambda$ is the cosmological constant, $l$ the corresponding curvature radius, and $d$ the number of spacetime dimensions, 
its conjugate is then defined as 
\begin{equation}
  V = - 8 \pi \frac{\partial M}{\partial \Lambda} \,.
\end{equation}
The modified first law of thermodynamics then reads 
\begin{equation}
  dM = \frac{\kappa}{2\pi} dA + V dP \,,
\end{equation}
plus possible black hole work terms as $\Phi dQ + \sum_i \Omega_i dJ_i$, which indicates that the mass of the black hole actually 
corresponds better to the enthalpy of an ordinary thermodynamic system, than to its internal energy. The mass of the black hole is 
thus equivalent to the amount of energy necessary to create the black hole and put it into its cosmological environment. 

The above mentioned articles have investigated various types of black holes with a cosmological constant, by analyzing the Gibbs 
free energy and its dependence on various thermodynamic quantities. For singly spinning black holes in $d$ dimensions, a transition 
between small and large black holes has been found, similar to the liquid-gas phase transition in fluids, and the corresponding 
critical exponents are the same as for a van der Waals fluid \cite{2012Kubi}. Analogous investigations have been done on singly 
and doubly spinning higher-dimensional AdS black holes in the canonical ensemble, i.e., fixed angular momentum (momenta). For the 
singly spinning ones, besides the usual transition between small and large black holes, there is a another reentrant phase transition 
from large back to small black holes, in analogy to the phase structure of multicomponent fluids \cite{2013Alta}. Ultimately, in 
multiply spinning black holes in $d=6$ dimensions, the thermodynamics of the system depends on the ratio $q = j_1 / J_2$ of the 
two angular momenta. For $q=0$, the system recovers the reentrant large/small/large phase transition structure of the singly spinning 
black hole. With nonzero, but small $q$, the phase transition changes to an analogy of a solid/liquid phase transition, whereas for 
$q \in [0.00905,0.0985]$, there is a complex phase transition structure with a change from large to intermediate and on to small 
black holes, in a phase diagram with two critical and one triple point. Ultimately, for $q > 0.0985$, a van der Waals type phase 
transition from liquid to gas is recovered \cite{2014Alta1}. 

With the introduction of the cosmological constant as an additional thermodynamic variable, it is thus possible to discover a 
very rich structure of thermodynamical properties and transitions in black holes, and by the analogy with fluids reinterpret some 
aspects of black hole thermodynamics \cite{2014Alta2}.

\section{Thermodynamic Structure Of The Space Of Equilibrium States Of The Kerr-Newman Black Hole Family}
In his talk, Miguel Angel Garcia-Ariza talked about a formulation of thermodynamics within a framework of differential geometry, 
taken in investigations in a collaboration with Merced Montesinos and Gerardo F. Torres del Castillo \cite{2014Garc,2015Garc}. 
In their work, they followed an approach to geometric thermodynamics defined by Ruppeiner \cite{rupp1979}, employing a metric 
formalism in the abstract $n$-dimensional manifold spanned by the extensive thermodynamic state variables, interpreted as 
coordinates on that manifold. Ruppeiner's metric is defined as the Hessian of the entropy function with respect to the extensive 
variables. The curvature defined by this metric is supposed to mirror the thermodynamic interactions of the system, i.e., a flat 
metric with zero curvature would correspond to a system without thermodynamic interactions, whereas some curvature implies that 
interactions are present, and curvature singilarities mark points of major changes in the system's properties. 

As an example of an simple hydrostatic thermodynamic system, the ideal gas was considered, with its fundamental equation given 
by 
\begin{equation}
  dU = T dS - p dV + \mu dN \label{eq:1} \,,
\end{equation}
with $U$, $T$, $S$, $p$, $V$, $\mu$, and $N$ denoting the standard thermodynamic state variables, i.e., the system's internal energy, 
temperature, entropy, pressure, volume, chemical potential, and number of particles, respectively. For systems following this general 
fundamental relation, the thermodynamic metric in equilibrium phase space can be written as 
\begin{equation}
  g_R = \frac{c_V}{T^2} dT^2 + \frac{1}{\kappa_T T V} dV^2 \,,
\end{equation}
where $c_V$ is the heat capacity at constant volume, and $\kappa_T$ is the compressibility at constant temperature. 

Since the ideal gas, having no interactions, is described to a flat metric in the manifold of equilibrium states, the question arises 
whether the ideal gas is the only and uniquely determined system with a flat thermodynamic metric. Assuming that for such systems, 
$c_V=\text{const}$, and employing some coordinate transformations, it can be shown that any thermodynamic system with a compressibility 
of the form 
\begin{equation}
  k_T^{-1} = e^t \left[tf_1 (V) +f_2(V) \right]^2 \,,
\end{equation}
where $t = \log T$, and $f_1,f_2$ are only functions of the volume, has a flat metric and vanishing curvature. As a consequence, the 
ideal gas is shown to represent only a particular case of a closed hydrostatic system with $C_V=\text{const}$. Such systems in general 
may have spaces of equilibrium states with a flat metric, despite the presence of interactions. This leads to some tension with the 
usual conjecture on the correspondence between curvature of the thermodynamic metric and interactions. An example was constructed, in 
which Ruppeiner's metric was flat, and yet the system featured thermdynamic interactions. 

A geometric formalism of thermodynamics can be applied also in the field of black hole thermodynamics, concretely for the example of 
the Kerr-Newman family of black holes. Some inconsistencies with the choice of the thermodynamic potential were pointed out, i.e., the 
dependence on the predicted critical points of the system on the thermodynamic potential used.

\section{Black Hole Thermodynamics In Finite Time}
A slightly 'engineering' point of view to black hole thermodynamics was provided in the contribution of Christine 
Gruber, who in collaboration with Alessandro Bravetti and Cesar Lopez-Monsalvo investigated the impact of finite-time 
effects on black hole engines \cite{2015Brav}. 

Finite-time thermodynamics is an approach to thermodynamics from a more realistic point of view, 
dropping the assumption of perfectly reversible processes and instead trying to estimate dissipative losses that occur 
along the evolution of a system in finite times -- so to speak, calculating the energetic 'price of haste'. 
Thermodynamic processes carried out in finite times suffer from dissipative losses because it is not possible to go 
through the path reversibly, i.e., as a sequence of perfect local equilibria. During a process which is not perfectly 
reversible, in each infinitesimal step along the way small amounts of energy are dissipated. Intuitively, this can be 
understood from a so-called horse-carrot process. To drive a system along a path in thermodynamic phase space, it is 
brought into contact with a much larger reservoir having slightly different values of the intensive thermodynamic 
quantities. Equilibration with a sequence of reservoirs causes the system to change the values of its intensive quantities 
in infinitesimal steps, eventually reaching the final point of the path. If there is however not an infinite time at hand 
to establish the perfect equilibrium in each step along the way, dissipative losses will occur and sum up over the length 
of the path. Quantitatively, these losses can be computed using a geometric formalism of thermodynamics \cite{rupp1979}, as 
was already introduced in the previous talk/chapter. 
A thermodynamic metric on the abstract manifold of thermodynamic phase space can be defined as the Hessian of the thermodynamic 
potential such as e.g. the entropy $S$, with respect to the extensive thermodynamic variables $x^{i}$ of the system, 
\begin{equation}
  ds^{2}= S_{ij} \,dx^i dx^j 
\end{equation}
From this definition, the length of a process from the initial to the final point can be defined in the thermodynamic phase 
space as 
\begin{equation}
  L_S = \int_{\gamma} \sqrt{-S_{ij} d{x^{i}} d{x^{j}}}\,.
\end{equation}
It has been shown \cite{SalamonBerryPRL,andresen2011current} that this length can be related to the sum (or integral) of 
the infinitesimal energy dissipations in each step along the process, 
\begin{equation} \label{SB1}
  (\Delta S)_{\rm diss}\geq L_{S}^{2}\,\frac{\epsilon}{\tau}\,,
\end{equation}
where $\epsilon$ is the infinitesimal amount of dissipation generated in each step along the path, and $\tau$ is the duration 
of the process. 

The aim of finite-time thermodynamics is to obtain realistic limitations on idealized scenarios, and it is thus a useful 
tool to reassess the efficiency of thermodynamic processes, with wide applications in industrial contexts. In the work 
by Gruber et. al, it has been applied to black holes in the context of Penrose-like processes, which consider the extraction 
of energy from a Kerr black hole. They investigated thermodynamic processes decreasing the angular momentum of the black hole 
from the extremal to the Schwarzschild limit, or the extraction of mass, and calculated the thermodynamic length of these 
processes. The results showed that the dissipative losses during the extraction of energy grow stronger close to the extremal 
limit, and thus in order to minimize dissipation, the extremal limit should be avoided.

\section{The Volume of Black Holes}
The contribution of Maulik Parikh was concerned with the definition of volume for black holes, or stationary 
spacetimes in general \cite{2006Maul}. The need for such a definition arises from the quantum gravitational argument 
that holography, or the encoding of information on the surface of a black hole instead of the bulk volume, leads to a 
radical decrease in entropy, since it is proportional to area$/l_p^2$, instead of volume$/l_p^3$. The problem here is 
however that this notion of volume is not well-defined, and depends on the interpretation of time and space across 
horizons, or on the choice of time slicing in a spacetime. Therefore, an invariant definition of volume was proposed as 
\begin{equation}
  V_{spatial} = \frac{dV_D}{dt} \,,
\end{equation}
where 
\begin{equation}
  dV_D (t) = \int_t^{t + dt} dt' \int d^{D-1}x \sqrt{-g_D} \,,
\end{equation}
which is the differential \emph{spacetime} volume of a given spacetime, and contains the determinant $g_{(D)}$ of the 
complete spacetime, instead of only the spatial part. This definition for the volume holds for spacetimes in which a 
timelike Killing vector exists (even if that Killing vector is not a global one), which is equivalent to the requirement 
of thermal equilibrium, and is thus valid for static or stationary spacetimes. The proposed expression for the volume 
does not change with time, and moreover does not depend on the choice of the stationary time-slicing. For a four-dimensional 
spherically symmetric spacetime, the volume takes on the form 
\begin{equation}
  V_{4d} = \frac{4\pi}{3} r_+^2 \,.
\end{equation}
With these definitions at hand, the possibility of having a black hole with a finite surface area, but an infinite volume 
was addressed. However, considering the symmetry groups of $3$- and $4$-dimensional spherical, flat and hyperbolic spacetimes, 
it can be argued that there is no class of solutions which admits an infinite volume that is bounded by a finite horizon area. 
The generalization of these arguments to higher dimensions remain to be done. 

In a second part of his talk, investigations on black hole nucleation have been presented \cite{2011Maul}. The rate $\Gamma$ 
of nucleation of a certain type of instantons in a chosen background is given by 
\begin{equation}
  \Gamma \simeq \frac{\mathrm{exp} \left( -I_E [\mathrm{instanton}] \right) }{\mathrm{exp} \left( -I_E [\mathrm{background}] \right) } \,,
\end{equation}
where $I_E$ are the Euclidean action. The nucleation rate of black holes in Einstein and Einstein-Gauss-Bonnet gravity for 
$4$-dimensional de Sitter spacetimes is calculated, resulting in 
\begin{equation}
  \Gamma \simeq \mathrm{exp} \left( -\frac{\pi L^2}{3G} + \frac{4\pi \alpha}{G} \right) \,,
\end{equation}
where $G$ is Newton's constant, $L$ is the curvature radius of de Sitter, and $\alpha$ is the Gauss-Bonnet coupling term. 
Thus, by adding a topological Gauss-Bonnet term to the gravitational action of four-dimensional de Sitter spacetime, the 
nucleation rate of black holes is greatly enhanced, which renders the theory very sensitive to instabilities. The Gauss-Bonnet 
coefficient thereby serves as a stability bound on the maximal curvature of spacetime.

\section{Entropy in locally-de Sitter spacetimes}
In her presentation, Adriana Victoria Araujo introduced work on the thermodynamic properties of de Sitter spacetimes, done in 
collaboration with J. G. Pereira \cite{2015Arau}. When constructing general relativity on a de Sitter background instead of a 
Minkowskian one, the local Riemannian geometry is modified, in particular spacetime is endowed with a de Sitter-Cartan connection, 
changing the local dynamics. As a consequence, the notion of entropy changes, which is directly related to Noether charges derived 
from the system's spacetime diffeomorphisms, which now are determined by the de Sitter group. The effects of the de Sitter background 
on the thermodynamics of black holes has been investigated. 

Due to the presence of a cosmological constant, or a curvature radius $l$, a Schwarzschild-de Sitter spacetime possesses two horizons. 
The usual Schwarzschild horizon is modified by terms including $l$, and given by 
\begin{equation}
  r_{SdS} = 2 M \Big(1 + \frac{4 M^2}{l^2} + \cdots \Big) \,,
  \label{HoriRadius}
\end{equation}
whereas the cosmological de Sitter horizon is 
\begin{equation}
  r'_{SdS} = l \Big( 1 - \frac{M}{l} - \frac{3 M^2}{2 l^2} + \cdots \Big) \,,
  \label{HoriRadius'}
\end{equation}
expanded in powers of $M/l$, assuming that the de Sitter curvature radius is much larger than the Schwarzschild radius. The corresponding 
thermodynamics of Schwarzschild-de Sitter spacetime is determined by both horizons and their properties. 

Considering the Schwarzschild horizon with de Sitter modifications, the entropy $S=A/4$ can be calculated as 
\begin{equation}
  S = 4 \pi M^2 \Big(1 + \frac{8 M^2}{l^2} + \cdots \Big) \,.
  \label{dSentro}
\end{equation}
Differentiating the entropy with respect to the mass, the temperature of the black hole horizon can be obtained as 
\begin{equation}
  T = \frac{1}{8 \pi M} \Big(1 - \frac{4 M^2}{l^2} + \cdots \Big) = \frac{\kappa}{2\pi} \,.
  \label{dStempe}
\end{equation}
which as well gives the surface gravity $\kappa$. Ultimately, the energy of the horizon is modified by the de Sitter curvature 
radius as well, and can be determined as 
\begin{equation}
  E = M + \frac{8 M^3}{l^2} + \cdots \, .
  \label{bhE}
\end{equation}
In the case of de Sitter horizon, the entropy is calculated as 
\begin{equation}
  S' = \pi l^2 \Big(1 - \frac{2 M}{l} - \frac{2 M^2}{l^2} + \frac{3 M^3}{l^3} + \cdots \Big) \,,
  \label{dSentro'}
\end{equation} 
with a corresponding horizon temperature of 
\begin{equation}
  T' \equiv \frac{1}{2 \pi r'_{SdS}} = \frac{1}{2 \pi l} \Big(1 + \frac{M}{l} + \frac{3 M^2}{2 l^2} + \cdots \Big)\,.
  \label{dStempe'}
\end{equation}
The modified energy of the de Sitter horizon then is 
\begin{equation}
E' = l - M - \frac{2 M^2}{l} + \frac{13 M^3}{12 l^2} \cdots \,. 
\label{dsE}
\end{equation}
In summary, the thermodynamic properties of the horizons in Schwarzschild-de Sitter spacetime have been investigated, and it was shown 
that the presence of a cosmological constant, or a finite de Sitter curvature radius, leads to a mutual dependence of the horizon properties.

\section{Conclusions}

The main conclusion derived from the talks presented in this parallel session is that nowadays black hole thermodynamics 
is a vast area of active research. The different geometric approaches are just one method which allows us to investigate 
the structure of the equilibrium and phase spaces, the thermodynamic volume of black holes, the phase transition structure, 
and the novel idea of finite-time black hole thermodynamics. Future works in this direction include the investigation of
the cosmological constant and other physical parameters as thermodynamic variables which completely change the structure of
the equilibrium and phase spaces. Also, the search for microscopic models for black holes is an open issue that can be
handled by using geometric methods.

However, other methods are necessary in order to attack the problem of understanding
the physical meaning of black hole thermodynamics. Entropy, which is perhaps the most important thermodynamic property of black holes, 
is far from being completely understood. A classical origin of entropy is a possibility, although a quantum origin is certainly
a very challenging idea. Both approaches were discussed in this parallel session. In addition, the idea of using black hole 
thermodynamics to detect deviations of generalized models from Einstein gravity is a very promising approach.

This parallel session was  held at a highly scientific level, and carried out in full cordiality. We thank all the speakers
and attendees for their contributions, discussions and suggestions that made possible this session.

\section*{Acknowledgments}
CG was supported by funding from the DFG Research Training Group 1620 `Models of Gravity', and by an UNAM postdoctoral fellowship 
program. O.L. wants to thank the National Research Foundation (NRF) for financial support and P. K. S. Dunsby for fruitful discussions. 
This work was partially supported by DGAPA-UNAM, Grant No. 113514, and CONACyT, Grant No. 166391.

\bibliographystyle{ws-procs975x65}
\bibliography{MG14}

\begin{thebibliography}{10}

\bibitem{frankel}
T.~Frankel, {\em The Geometry of Physics: An Introduction} (Cambridge
  University Press, Cambridge, UK, 1997).

\bibitem{ym53}
C.~N. Yang and R.~L. Mills, {\em Phys. Rev.}   (1954).

\bibitem{callen}
H.~B. Callen, {\em Thermodynamics and an Introduction to Thermostatics} (John
  Wiley \& Sons, Inc., New York, 1985).

\bibitem{huang}
K.~Huang, {\em Statistical Mechanics} (John Wiley \& Sons, Inc., New York,
  1987).

\bibitem{rao45}
C.~R. Rao, {\em Bull. Calcutta Math. Soc.} {\bf 37}  (1945).

\bibitem{amari85}
S.~Amari, {\em Differential-Geometrical Methods in Statistics} (Springer,
  Berlin, 1985).

\bibitem{wei75}
F.~Weinhold, {\em J. Chem. Phys.} {\bf 63}, 2479, 2484, 2488, 2496  (1975).

\bibitem{wei75-2}
F.~Weinhold, {\em J. Chem. Phys.} {\bf 65}, p. 558  (1975).

\bibitem{rupp1979}
G.~Ruppeiner, {\em Phys. Rev. A} {\bf 20}, 1608  (1979).

\bibitem{rup95}
G.~Ruppeiner, {\em Rev. Mod. Phys.} {\bf 67}, p. 605  (1995).

\bibitem{rup96}
G.~Ruppeiner, {\em Rev. Mod. Phys.} {\bf 68}, p. 313  (1996).

\bibitem{her73}
R.~Hermann, {\em Geometry, physics and systems} (Marcel Dekker, New York,
  1973).

\bibitem{arnold}
V.~I. Arnold, {\em Mathematical Methods of Classical Mechanics} (Springer, New
  York, 1980).

\bibitem{quev07}
H.~Quevedo, {\em J. Math. Phys.} {\bf 48}  (2007).

\bibitem{2013Abdo}
S.~Abdolrahimi, C.~Cattoen, D.~N. Page and S.~Yaghoobpour-Tari, {\em JCAP} {\bf
  06}, 039  (2013).

\bibitem{1997Frol}
V.~Frolov and I.~Novikov, {\em Black Hole Physics: Basic Concepts and New
  Developments} (Kluwer Academic Publishers, 1997).

\bibitem{2012Alex}
S.~Alexeyev and K.~Rannu, {\em JETP} {\bf 114}, 406  (2012).

\bibitem{2011Cucc}
A.~Cucchiara, A.~Levan and D.~F. et~al., {\em Astrophys. J.} {\bf 736}, 7
  (2011).

\bibitem{2015Alex}
S.~Alexeyev, K.~Rannu, P.~Dyadina, B.~Latosh and S.~Turyshev, {\em JETP} {\bf
  147}, 1120  (2015).

\bibitem{1974Hawk}
S.~W. Hawking, {\em Nature} {\bf 248}, 30  (1974).

\bibitem{2015Acqu}
G.~Acquaviva, G.~F.~R. Ellis, R.~Goswami and A.~I.~M. Hamid, {\em Phys. Rev. D}
  {\bf 91}, 064017  (2015).

\bibitem{1958Bel}
L.~Bel, {\em Compt. Rend. Acad. Sci.} {\bf 247}, 1094  (1958).

\bibitem{2013Mahj}
A.~Majhi, {\em arXiv:1303.4832 [gr-qc]}   (2013).

\bibitem{2012Kubi}
D.~Kubiznak and R.~B. Mann, {\em JHEP} {\bf 07}, 033  (2012).

\bibitem{2013Alta}
N.~Altamirano, D.~Kubiznak and R.~B. Mann, {\em Phys. Rev. D} {\bf 88}, 101520
  (2013).

\bibitem{2014Alta1}
N.~Altamirano, D.~Kubiznak, R.~B. Mann and Z.~Sherkatghanad, {\em Galaxies}
  {\bf 2}, 89  (2014).

\bibitem{2014Alta2}
N.~Altamirano, D.~Kubiznak, R.~B. Mann and Z.~Sherkatghanad, {\em Class. Quant.
  Grav.} {\bf 31}, 4  (2014).

\bibitem{2015Kubi}
D.~Kubiznak and R.~B. Mann, {\em Can. Journ. Phys.} {\bf 93 (9)}, 999  (2015).

\bibitem{2014Garc}
M.~A. Garcia-Ariza, M.~Montesinos and G.~F.~T. del Castillo, {\em Entropy} {\bf
  16}, 6515  (2014).

\bibitem{2015Garc}
M.~A. Garcia-Ariza, {\em arXiv:1503.00689 [math-ph]}   (2015).

\bibitem{2015Brav}
A.~Bravetti, C.~Gruber and C.~S. Lopez-Monsalvo, {\em arXiv:1511.06801 [gr-qc]}
  .

\bibitem{SalamonBerryPRL}
P.~Salamon and R.~S. Berry, {\em Physical Review Letters} {\bf 51}, 1127
  (1983).

\bibitem{andresen2011current}
B.~Andresen, {\em Angewandte Chemie International Edition} {\bf 50}, 2690
  (2011).

\bibitem{2006Maul}
M.~Parikh, {\em Phys. Rev. D} {\bf 73}, 124021  (2006).

\bibitem{2011Maul}
M.~Parikh, {\em Phys. Rev. D} {\bf 84}, 044048  (2011).

\bibitem{2015Arau}
A.~V. Araujo and J.~G. Pereira, {\em Int. J. Mod. Phys. D} {\bf 24}, 1550099
  (2015).

\end{thebibliography}

\end{document}